\documentclass[pdflatex,sn-mathphys-num]{sn-jnl}

\usepackage{geometry}
\usepackage{graphicx}%
\usepackage{multirow}%
\usepackage{amsmath,amssymb,amsfonts}%
\usepackage{amsthm}%
\usepackage{mathrsfs}%
\usepackage[title]{appendix}%
\usepackage{xcolor}%
\usepackage{textcomp}%
\usepackage{manyfoot}%
\usepackage{booktabs}%
\usepackage{algorithm}%
\usepackage{algorithmicx}%
\usepackage{algpseudocode}%
\usepackage{listings}%
\usepackage{url}

\begin{document}

\title[The Economics of an Open-Source Quantum Computer]{The Economics of an Open-Source Quantum Computer}


\author*[1,2,5]{\fnm{Francesco} \sur{Bova}}\email{francesco.bova@Rotman.utoronto.ca}

\author[2,3,4,5]{\fnm{Roger G.} \sur{Melko}} 

\affil*[1]{\orgdiv{Rotman School of Management}, \orgname{University of Toronto}, \city{Toronto}, \state{Ontario}, \country{Canada}}

\affil[2]{\orgdiv{Creative Destruction Lab}, \city{Toronto}, \state{Ontario}, \country{Canada}}

\affil[3]{\orgdiv{Dept. of Physics \& Astronomy}, \orgname{University of Waterloo}, \city{Waterloo},  \state{Ontario}, \country{Canada}}

\affil[4]{\orgdiv{Perimeter Institute for Theoretical Physics}, \city{Waterloo},  \state{Ontario}, \country{Canada}}

\affil[5]{\orgdiv{Open Quantum Design}, \city{Waterloo},  \state{Ontario}, \country{Canada}}


\abstract{Open-source projects that aim to make their offerings public have competed against for-profit, proprietary companies in a number of domains. These open-source projects often arise in response to the offerings of proprietary companies in markets where products have already been commercialized. We assess what impact an open-source project might have when it enters the market for quantum computing -- a market where the core technology is still being developed. We argue that an open-source quantum computer might alleviate market frictions that have impeded the development of a fault-tolerant quantum computer by providing the market with a possible mechanism for: 1.) benchmarking, 2.) the development of hardware agnostic technology, and 3.) improved liquidity on the supply side of the market for labor. Should these outcomes be realized, they may not only benefit the open-source project, but also the for-profit, proprietary companies operating in the ecosystem. In this respect, an open-source quantum project may play a more complementary role to the proprietary firms in the ecosystem, rather than a more competitive role. Our collective insights may have implications for other settings where an open-source project enters into a market where the core technology has yet to be commercialized.}




\maketitle

\section{Introduction}\label{sec1}

The open-source model is a well-known approach to creating a product offering in the software and semiconductor industries \cite{OpenMITReview}. At their heart, open-source models aim to make their offerings (e.g., source code, specifications, designs, etc.) public.  These public offerings can then be used, downloaded, studied, modified, and redistributed with modifications – typically for free. In the case where the open-source offerings are hardware design specifications, toolchains, etc., the public is permitted to use these specifications to build their own version of the open-source hardware. 

There are many markets where open-source initiatives compete with proprietary, for-profit companies. These markets include: the market for operating systems where, for example, open-sourced Linux competes against proprietary Microsoft Windows; the market for internet browsers where, for example, open-sourced Mozilla competed against proprietary offerings like Internet Explorer; and the market for microchip manufacturing and microchip development where RISC-V competes against proprietary players like ARM. The economics literature has provided extensive insights into the market implications of including an open-source competitor in a market made up of proprietary, for-profit companies. Common findings include the results that an open-source competitor: 1.) puts price pressure on incumbent proprietary firms (because open-source offerings are typically free) and 2.) often affects the quality of the proprietary technology in equilibrium (See, for example, discussions in \cite{Bitzer},\cite{economo},\cite{sen}). 
Additional findings suggest that open-source offerings can: 1.) be used as a catalyst for market creation \cite{Fitzgerald}, 2.) mitigate risk in the development of new technologies (e.g., evidence suggests that vulnerabilities are often found more quickly in open-source software projects than in proprietary software projects) \cite{Ebert}, and 3.) be more challenging to work with for end users because they often offer fewer support services or have less user-friendly interfaces than their proprietary peers \cite{sen}.

Most open-source projects are initiated in markets where proprietary firms have already commercialized their offerings.\footnote{A notable exception is the internet itself. Many technologies in the internet's stack, such as the Apache HTTP server, originated from open-source development before the internet was widely commercialized.} This outcome occurs because open-source projects are often created in response to products that: 1.) have already been commercialized and are being sold to consumers, and 2.) have characteristics that the founders of the open-source project felt could be improved upon. For example, the early GNU operating system, an open-source project created by Richard Stallman, was created partially in response to the buggy software that was being sold by for-profit firms at the time \cite{Chesbrough}.  Less common are settings where an open-source project enters a market that is still in the R\&D phase of the technology cycle. These markets are characterized by firms that are research intensive, often pre-revenue, and with offerings that have yet to be commercialized. There are presumably few open-source projects in these markets because there is no incumbent product that an open-source project's offering could potentially improve upon. 

We discuss the impact that an open-source project might have on a market that is still in its R\&D stage by focusing on a specific nascent technology that has yet to be commercialized – quantum computing. 
Quantum computers, if scaled up significantly from today's laboratory devices, have the potential to vastly speed up a number of challenging calculations. 
This makes quantum computers a potentially transformative technology, although a significant amount of research and development is still required in order to achieve this outcome.
The market for quantum computing is fairly novel relative to other deep tech markets because we have yet to build a sufficiently large and accurate quantum computer that can solve a commercially valuable problem in a significantly timelier manner than a classical computer. As a result, all players in the market are still in the process of developing the technology. 

While an open-source quantum project may compete directly with proprietary quantum computing companies once the technology matures, we argue that it may also mitigate market frictions that currently impede the development of the technology. We argue that the alleviation of these market frictions may reduce the time it takes to create a fault-tolerant quantum computer for not only the open-source entrant, but for all companies in the ecosystem. In turn, an open-source quantum computing project may play a more complementary role in the market for quantum computing as opposed to the more competitive role that we often associate open-source projects with. We believe our analysis has implications for other settings where an open-source project enters a market where the core technology has yet to be fully developed. We expand on these ideas below.

\section{The market for quantum computing}\label{sec2}

Quantum computers differ from the conventional computers that we use every day in a number of ways. The foundation of a quantum computer is the qubit which, when combined with an appropriate control and software stack, will allow certain algorithms to be executed in novel ways. One central technical challenge in building a quantum computer is achieving {\it fault tolerance}. Fault tolerance arises when computations can continue to be executed correctly even in the presence of errors in the qubit and control infrastructure. 
Should a sufficiently large, fault-tolerant quantum computer be built, there is optimism that it will be able to solve certain intractable problems, such as the important problem of prime number factorization, in a timelier manner than its classical counterparts \cite{Shor}.
Despite their extraordinary promise, we have yet to develop a fault-tolerant quantum computer that can solve a problem of commercial value. 
Recent work, like Google's error correction demonstration \cite{Acharya2024}, are an important step in the right direction, however significant R\&D work remains before true fault tolerance is achieved.

There are a growing number of proprietary, for-profit companies that are currently attempting to build a full-stack, fault-tolerant quantum computer. These companies are exploring different architectures at each layer of the quantum stack to achieve this goal. The market is currently made up of aspiring startups, several publicly-traded companies, 
and several large well-capitalized technology firms that are attempting to build a quantum computer as part of a broader suite of technology offerings. However, as we note above, whether as a stand-alone firm or as a business segment within a larger company, none of these initiatives has yet to achieve large-scale fault tolerance. Moreover, the timeline to achieve the elusive goal of large-scale fault tolerance continues to be uncertain.

\begin{figure}[t]
\centering
\includegraphics[width=0.75\textwidth]{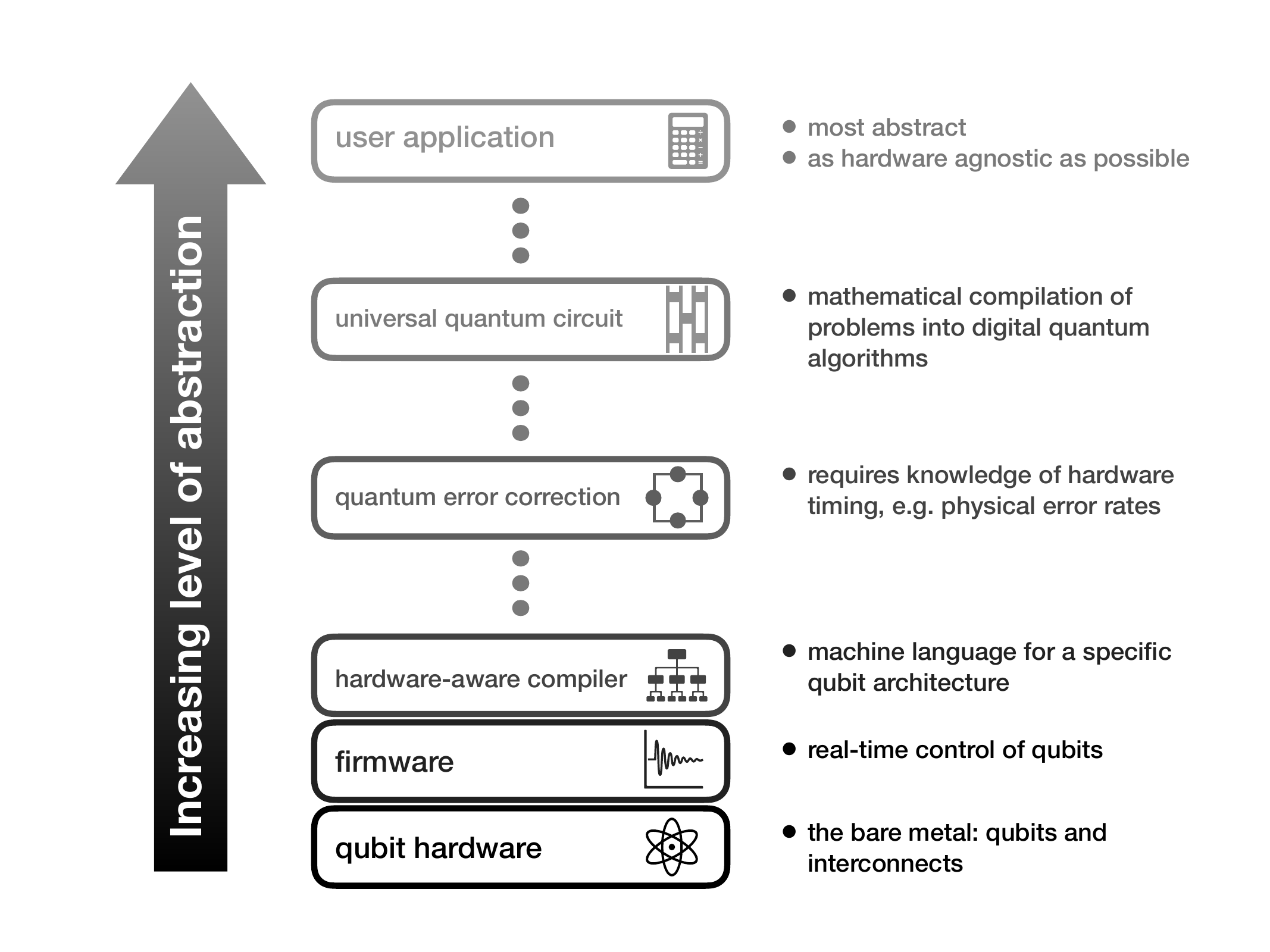}
\caption{The quantum computing stack.  The base of the stack is the specific qubit hardware.  As one moves up the stack, the level of abstraction generally increases, until layers become ``hardware agnostic’’.}\label{fig1}
\end{figure}

A quantum computer is comprised of a bundle of technologies, generally referred to as layers of a ``stack'' (see Figure~\ref{fig1}), where varying degrees of abstraction can exist between each layer \cite{Open_standards_stack}.  Experts have suggested different approaches to creating the various layers. For example, the ``bare metal'' or qubit layer is the hardware base of the stack. It includes the physical qubits and their interconnects which will be used to power the information processing of the quantum computer.  Bare metal designs are essentially never open-sourced, except in academia. Above the bare metal layer is the real-time control, which we call the ``firmware'' layer.  Utilizing conventional on-chip integrated circuits, such as field-programmable gate arrays (FPGAs), this layer is the interface between the quantum hardware and the classical software required to run it.  This layer can include features like on-chip control, time-dependent pulse generation, and hardware calibration.  Proprietary hardware systems usually have proprietary firmware (with some exceptions \cite{OpenHardware}). Above the firmware layer is the hardware aware compiler.  This layer takes instructions from an intermediate representation (e.g., a quantum state preparation and measurement protocol that may be fairly abstract) and converts it to a set of instructions that will control the firmware.  This level of compiler needs to know quite a lot about the hardware, hence it is generally proprietary in computers where the bare metal design is proprietary. Continuing up the stack, we have intermediate representations, and higher levels of abstraction.  Some of these layers (for example, quantum error correcting protocols) still need a high level of hardware awareness \cite{Silva}.  Others, like a compiler of abstract digital circuits to logical qubits and gates, can be less hardware aware.  Generally, the requirement for hardware awareness decreases as one moves up the stack, and the possibilities for independence between the layers increases. As a result, there is hope that, at least for some of the layers in a quantum computer, a more hardware agnostic approach may be possible. Finally, note that at every layer, firms make an initial choice as to the approach they would like to take. For example, at the bare metal layer, several different architectures can feasibly be used to create qubits, such as superconducting circuits, neutral atoms, trapped ions, quantum dots, etc.
This choice is somewhat analogous to the choice made in the semiconductor industry in the 1950s and 1960s between germanium and silicon \cite{SilGer}. 
Similar to the approaches available at the bare metal layer, numerous approaches are possible at each layer of the quantum computing stack.

\section{Why is building a quantum computer so expensive?}\label{sec3}

\begin{figure}[t]
\centering
\includegraphics[width=0.7\textwidth]{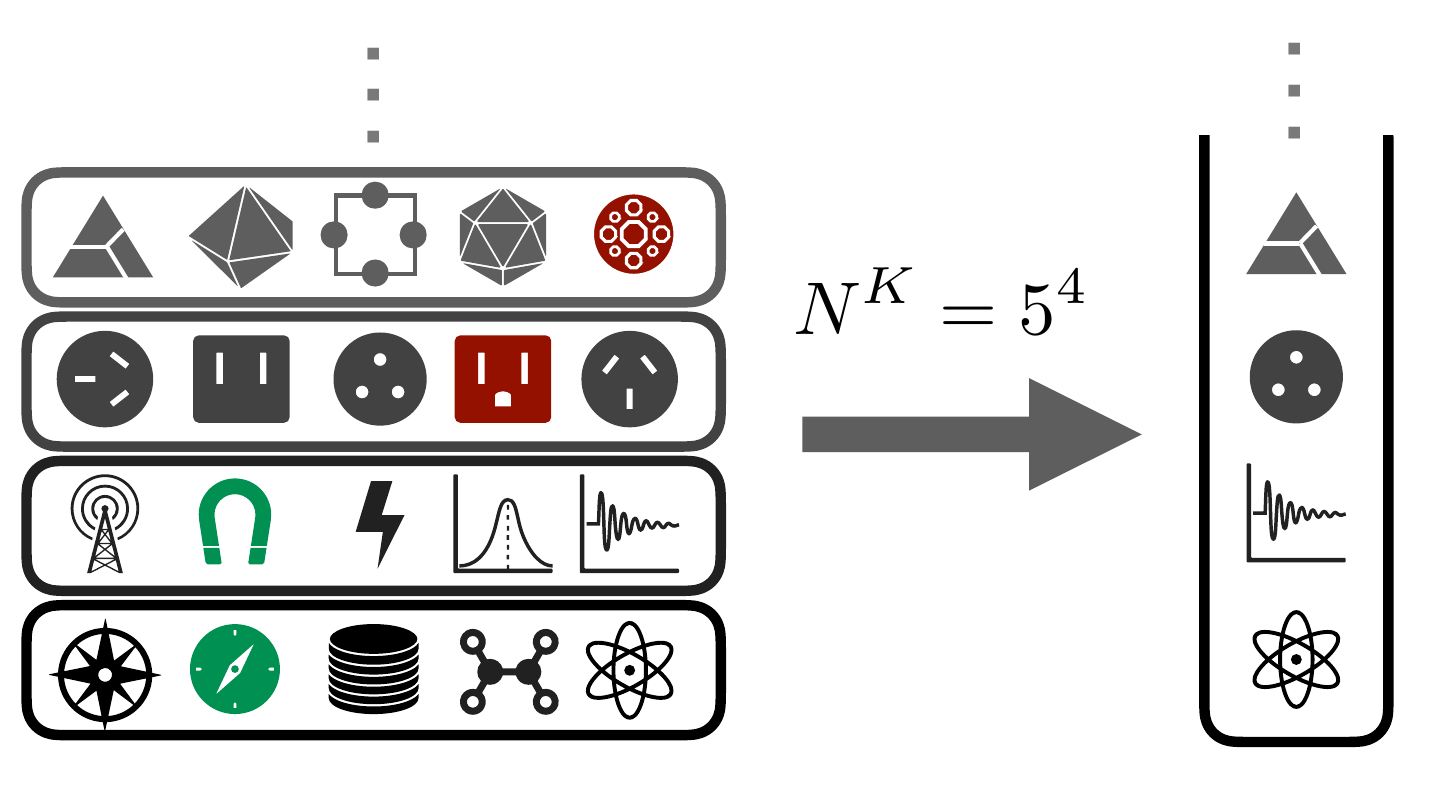}
\caption{The curse of dimensionality is illustrated above where the possible technologies for multiple layers of the stack (left) must be reduced to make one choice (right) for the total architecture of a full-stack quantum computer.
In this example, there are $K=4$ layers of the stack, and there are $N=5$ approaches to developing each layer. If any approach in a given layer could be selected independently of any other approach, there would be $N^K = 5^4 = 625$ possible configurations to developing a fault-tolerant quantum computer. The curse of dimensionality can be partially mitigated by correlations between the value to using certain technologies together within the stack. The figure above provides examples of technological complements (in green) and substitutes (in red) across layers in the stack.}\label{fig2}
\end{figure}

Building a full-stack quantum computer is perhaps the most challenging technological task ever embarked upon.
While it is not clear how many approaches there are or will be to building each layer of a full stack quantum computer, as a thought experiment we can conduct a quick back-of-the-envelope calculation. 
Consider the goal of creating a single technology bundle -- a quantum computer -- out of a stack with $K$ layers, where there are $N$ possible approaches to developing each layer of the stack.
A very simple approximation would be to assume that an approach to any layer in the stack can be chosen independently of an approach to any other layer in the stack. This would result in $N^K$ possibilities for the bundle (Figure \ref{fig2}).
For example, imagine that there are $N=10$ possible approaches per layer to creating $K=5$ layers of a full-stack quantum computer. In this example, there would be $10 \times 10 \times 10 \times 10 \times 10 = 10^5 = 100,000$ possible configurations to building a full-stack quantum computer. 
The hope for any proprietary quantum computing company would be that at least one of these full stack approaches will eventually lead to the creation of a fault-tolerant computer that can solve a problem of commercial value (see Figure~\ref{fig2}).
Notice that, in this model, the number of possible configurations for a full-stack quantum computer grows exponentially with every possible approach for a given layer. This type of exponential increase is frequently referred to as the curse of dimensionality, as it implies an exponential increase in the number of configurations that need to be searched through to find a suitable configuration. 

We note that there are factors that may both inherently mitigate or exacerbate this curse of dimensionality. First, the choices around which technology to use in each layer of the stack are typically not independent (i.e., the value of using one approach for a certain layer of the stack is often correlated with the value of using another approach for a different layer of the stack). Evidence suggests that if there are complementary or substitutionary relationships among technologies in a technology bundle then the curse of dimensionality may be mitigated, since the propensity for firms to make choices between different technologies in a bundle is not independent \cite{YU2012354}. Complementary relationships arise when the value of using one technology is positively correlated with the use of another, and substitutionary relationships arise when the value of using one technology is negatively correlated with the value of using another (see Figure~\ref{fig2}).

 These sorts of correlated relationships often arise between various technologies in the layers of a quantum computer stack. For example, when atomic qubits are selected as the bare metal layer, optical (laser) systems must be used for qubit control which impacts the firmware layer. Thus, an optical pulse control layer is perfectly complementary to the atomic qubit layer at the bare metal level because an atomic qubit layer will not function without optical control. Building on our numerical example, it follows that, rather than there being $1 \times 10 \times 10 \times 10 \times 10 = 10^4 = 10,000$ possible configurations to building a quantum computer conditional on selecting atomic qubits at the bare metal layer, there would only be $1 \times 1 \times 10 \times 10 \times 10 = 10^3 = 1,000$ possible configurations.  In this example, the complementary relationship between these two technologies within the stack will reduce the possible configurations to building a quantum computer, and in turn help to mitigate the curse of dimensionality.

Relatedly, there are various technologies in the stack that act as substitutes to one another. Technologies are substitutes when using them together in the stack is less desirable or simply not possible. Thus, the likelihood of observing them being used together is negatively correlated. For example, superconducting qubits at the bare metal layer cannot be combined with optical control technology at the firmware layer. So, any configuration that makes use of both superconducting qubits and optical control will lead to a configuration to building a fault-tolerant quantum computer that is not viable. It follows that this combination of technologies would reduce the number of possible configurations to creating a quantum computer in our numerical example by $1 \times 1 \times 10 \times 10 \times 10 = 10^3 = 1,000$, which in turn would also mitigate the curse of dimensionality.

Importantly, while correlations between the value of using various technologies within the stack may mitigate the curse of dimensionality, other factors may exacerbate it. For example, our simple example arbitrarily models ten possible choices that firms can make at each layer of the stack. These choices may be underrepresented because there may be new and innovative approaches to various layers of the stack that have yet to be discovered. For example, arrays of neutral atoms were once not considered to be contenders for scalable qubit architectures.  However, breakthroughs in optical tweezers \cite{Ashkin} and related control systems have led to the increased competitiveness of quantum computers based on neutral Rydberg atoms at the bare metal layer \cite{Browaeys2020,Bluvstein2024}. Presumably new approaches will be discovered at every layer of the stack in the years to come. It is unclear what the upper bound might be for different approaches to different layers of the stack, but the greater the number of approaches the more severe the curse of dimensionality. Taken together, while there are factors that may mitigate or exacerbate the dimensionality of developing a quantum computer, dimensionality will still be a material factor in the development of quantum computers. This dimensionality can become a ``curse'' for firms if the cost to developing any particular configuration is comparatively expensive.

Unfortunately, ``expensive'' is an appropriate adjective to describe the cost to developing a hardware or software approach at almost any layer of the quantum computing stack. These costs result from the many challenges to creating a quantum computer that were comparatively inexpensive to address when the first classical, silicon-based computers were created. These challenges include quantum computers’ extreme sensitivity to noise, obstacles to scaling the hardware (such as the need to use cryogenics for cooling), uncertainty around approaches to error correction, and others. These headwinds have made the R\&D process very experimental and, in some cases, very expensive.

Of course, the large costs to developing a quantum computer need to be funded somehow. As the technology has yet to be commercialized, there are comparatively few avenues for quantum computing companies to generate revenue to offset R\&D expenses.\footnote{Companies building a full-stack quantum computer currently have imperfect machines which have been used to generate revenue from paid pilots, consulting, and other areas where market stakeholders are experimenting with the technology. In general, however, the cash inflows from these sources of revenue do not do enough to offset the cash outflows related to research and development.}  
As a result, large amounts of external financing, which are frequently sourced from capital markets, have been needed to fund the R\&D process. This outcome has led to a tendency for firms to be highly proprietary with the technologies they develop, because providers of capital expect to generate a substantial return on their significant investments. This need to generate a substantial return has also meant that hardware agnostic approaches to developing any particular layer, which could theoretically be useful to all firms in the ecosystem, have been generally avoided in favor of approaches that are linked back specifically to the proprietary hardware being developed in other parts of the stack \cite{Fingerhuth}. From a capital provider's perspective, this approach may make sense. After all, why make significant investments in a firm trying to commercialize a nascent technology if all of the firm’s competitors get to share in the returns?

With these points as a backdrop, we argue that it is not only the significant costs to building a quantum computer that have made firms in the ecosystem highly proprietary, but also the highly proprietary nature of the market that has, in turn, exacerbated the costs to building a fault quantum computer. For example, it can be challenging for a quantum computing firm to publicly illustrate the efficacy of its approach in any given layer via third-party benchmarking, because doing so may reveal proprietary information to the market. Relatedly, startups have created stand-alone businesses in an attempt to develop and commercialize technologies for certain layers of the stack (for example, around qubit noise models or error correction strategies). Notably, these ventures are not also building a full-stack quantum computer. As a result, these companies frequently do not have their own hardware to test on for the purpose of benchmarking. Rather, these companies need to benchmark outside of the lab by working with proprietary quantum computing companies that are trying to build a full-stack quantum computer. However, these full-stack companies may be reluctant to work with external solutions providers because the joint work might again reveal proprietary information \cite{Fingerhuth}. Separately, proprietary quantum computing firms may be reluctant to work with a stand-alone company if the company’s technology is too close to the IP that the full-stack company is already developing. In these settings, there may be a greater risk that one firm may inadvertently give away proprietary or strategic knowledge to the other \cite{Paasi}. Taken together, it may be challenging to benchmark the efficacy of enabling technologies produced by firms that are also not building a full-stack quantum computer against a common standard, because the costs to a for-profit quantum computing company to potentially revealing proprietary information may outweigh the benefits to potentially reducing its own development costs. The lack of a mechanism for effective benchmarking for these firms may stifle the development of new technologies in general, and presumably extends the path to fault tolerance and commercialization for the ecosystem as a whole.

The proprietary nature of the ecosystem has also led to a culture of strong NDAs for staff and employees, meaning that often, only those stakeholders who work for firms are active in developing the technology.  This outcome stifles the development of the technology more generally because external scientists often have a limited ability to improve on the technology that is currently under development.  Relatedly, the demand for talent with the requisite quantum background in the market for labor has historically outstripped the supply \cite{McKinsey}.  Conventional economics would suggest that the cost of labor should be comparatively high in such a setting, and comparatively high labor costs further inflate the cost base of a full-stack quantum computing company. The upskilling of quantum-adjacent employees (i.e., employees with, for example, programming skills but limited quantum background) has been suggested as a potential soluton to help address this gap between supply and demand \cite{McKinsey}.  However, upskilling new employees in this space is also non-trivially challenging. First, there is a steep learning curve for new developers when learning to work with quantum computing companies \cite{Fingerhuth}. Second, market stakeholders often do not have access to state-of-the-art quantum infrastructure, presumably because of the proprietary nature of the current ecosystem \cite{McKinsey24}.  This makes it challenging for prospective quantum-adjacent employees to upskill their abilities in the domain. These economic frictions create a barrier for prospective quantum and quantum-adjacent employees to entering the quantum ecosystem, and exacerbate the gap between demand and supply in the market for labor. The greater this gap, the higher the cost of labor and presumably, the higher the cost base for companies that are attempting to build a quantum computer.

Taken together, the substantial costs to creating a full-stack quantum computer have led firms to be highly proprietary in their approach to IP. Paradoxically, these highly proprietary approaches may have, in turn, increased each firm’s already substantial cost structures. The aggregate affect of these outcomes is that firms can often get “locked” into their approaches at different layers of the stack because it is too costly to diversify their approaches. On a related note, while complements and substitutes within the stack may reduce the curse of dimensionality to developing a quantum computer, they may also exacerbate the costs to being locked-in to a certain approach. This outcome arises because an approach to a layer that is complementary in one full-stack configuration may be substitutionary in another configuration. The costs to being locked-in to a certain path are potentially non-trivial, and can be particularly expensive when an approach to developing a full-stack quantum computer ends in failure.  For example, Microsoft invested in a type of qubit at the bare metal layer called a Majorana fermion that they hoped would significantly reduce error correction overhead if it could be harnessed for quantum computing.  However, after years of trying to synthesize one in the lab, the Majorana particle has never been convincingly demonstrated \cite{Majorana}. Microsoft's quantum computing ambitions were hence set back. This setback was potentially made more significant because this chosen bare metal layer was complementary to technologies being developed in other layers of its stack. 

\section{How Might an Open-Source Quantum Computer Project Alleviate These Problems?}\label{sec4}

For our purposes, an open-source quantum computer is one where the hardware, firmware, and software designs, standards, and codes are made freely available, and may be redistributed and modified  \cite{OpenHardware}.  In particular, design information provided about the quantum computer's hardware is easily discerned so that others can manufacture it (e.g., providing CAD drawings, bills of material, assembly instructions, qubit error models, etc.).\footnote{In reality, an open-source quantum project may face economic frictions which lead it to not make all of the IP public or to restrict participation on some dimension. Nevertheless, we consider it a valuable starting point to assume that all of the IP in our example will be open and that any market stakeholder can participate in the project.}  We discuss how such a computer might impact the quantum computing ecosystem given that the industry is currently in the R\&D phase of the technology cycle.

\subsection{As a catalyst to create hardware agnostic technology}

\begin{figure}[t]
\centering
\includegraphics[width=0.85\textwidth]{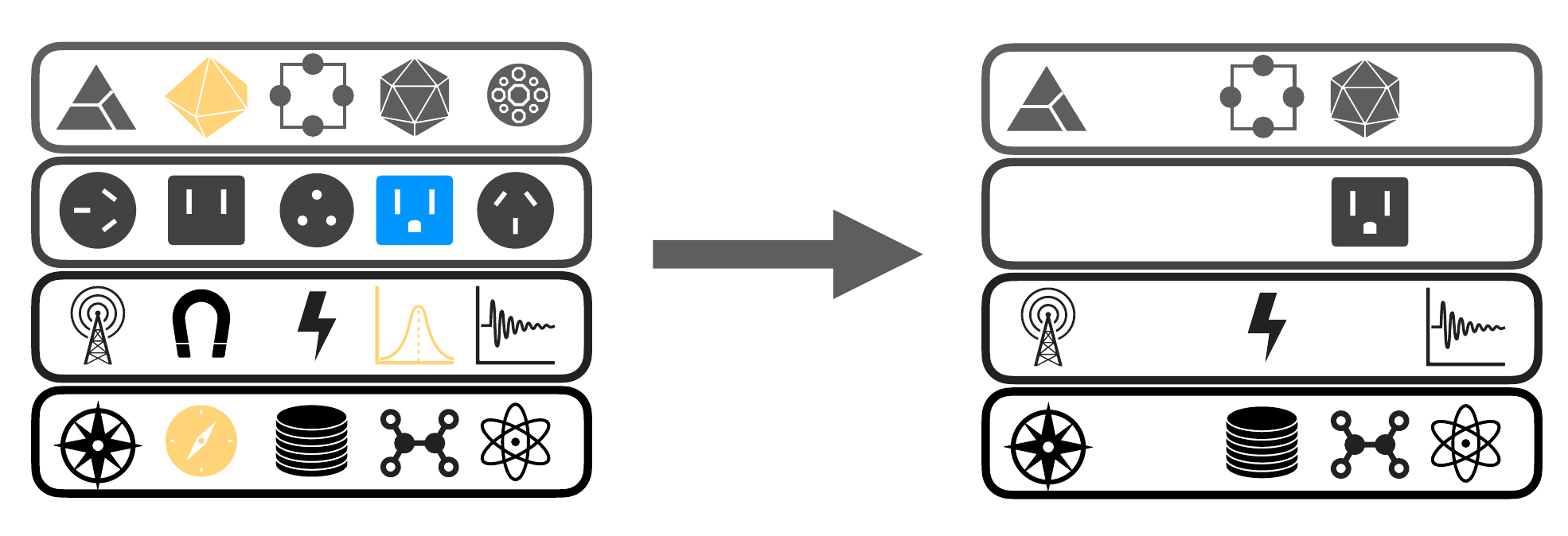}
\caption{The large number of choices at each layer of the stack might be reduced with mechanisms that increase exploration or reduce possible options.
On the left, blue approaches represent hardware agnostic technologies created by open-source providers that are subsequently adopted by the quantum ecosystem, while yellow technologies are those that have been eliminated through open-source community benchmarking.
On the right, the remaining technological choices are reduced further, by taking into account the complementary and substitutionary relationships within the stack that we highlight in Figure \ref{fig2}.}\label{fig3}
\end{figure}

 In our setting, the development of an open-source quantum computer will introduce a player into the ecosystem with technology that is not proprietary. This organizational choice may have several direct and indirect effects on the ecosystem. Like many open-source projects, market participants would be able to work on the open-source design and develop interesting new approaches for different layers in the stack. With access to, and input from, a variety of market stakeholders, an open-source project may eventually lead to the development of a hardware agnostic approach to a layer of a full-stack quantum computer. If a hardware agnostic approach can be created that is usable by all stakeholders in the ecosystem, then presumably the ecosysten as a whole will benefit. 
 For example, if a hardware agnostic approach can be developed at, say, the compiler layer, then that would be one layer of the stack that presumably would not have to be developed by other proprietary companies in the ecosystem. Building on our previous numerical example, such an outcome would reduce the number of possible configurations to work on, as there would only be four remaining layers that would need to be developed. This outcome would reduce the number of possible full-stack configurations from 100,000 to $10 \times 10 \times 10 \times 10 \times 1 = 10^4 = 10,000$. With fewer possible configurations, quantum computing companies would have a shorter path to building a fault tolerant quantum computer in expectation, as illustrated in Figure~\ref{fig3}. 
 This outcome should also reduce the expected costs to building a fault tolerant quantum computer.

\subsection{As a mechanism to improve benchmarking}

Utilizing an open-source project as a mechanism for benchmarking is not a new concept. For example, Imagenet, an open-source project which created a database of millions of images \cite{Imagenet}, holds a yearly contest for competing software platforms to accurately classify images from the database. This contest allows competing platforms to benchmark themselves against one another, and has led to some notable result’s including Alexnet’s highly-publicized accomplishment in 2012 where it achieved an error rate that was more than 10\% lower than the second-place finisher \cite{AlexNet}.  Importantly, the Imagenet project has had a significant impact on improving the efficacy of AI technology in the computer vision ecosystem, with alumni from the competition now represented across startups, tech giants, and academic institutions \cite{Quartz}.  Notably, the commercialization of computer vision technology predates the creation of Imagenet, with early efforts starting in the 1970s and the commercialization of facial recognition technology beginning in the 1990s \cite{Cvision}.  Thus, while Imagenet helped to improve the efficacy of the technology more generally via its benchmarking contests, it did so after the technology had already been commercialized and the industry was already somewhat mature. An open question is whether similar benchmarking outcomes can be achieved in a market that has yet to commercialize its core technology.

As we noted earlier, it can be challenging for market participants to benchmark the efficacy of a given approach to a certain layer of the quantum stack because: 1.) the ecosystem tends to be proprietary, and 2.) most stakeholders do not have access to the inner workings of a quantum computer. An open quantum project may allow all individuals, firms, institutions, etc., the ability to publicly benchmark their technologies on the same architecture. A simple way to envision this outcome is to define a metric, and to measure that metric in a very transparent way on an open quantum device \cite{MetriQ}. For example, the quantum volume of an open quantum stack could be compared to the quantum volume of a proprietary stack. In this case, even if the open-source computer does not have the “best” quantum volume, the procedure by which it is defined and measured can be perfectly transparent. 

This ability to transparently benchmark may have an impact on a variety of market outcomes. First, it may hasten the development of vital enabling technology which may benefit the ecosystem as a whole. Second, and perhaps of equal importance, it may provide public insights into approaches that should and should not be pursued. For example, if an open-source project can provide early indications that a certain approach to the bare metal layer generates a quantum volume that is comparatively low, then these results may inform the technology roadmaps of the rest of the ecosystem. Taken together, an open-source quantum project may result in a mechanism for benchmarking which will allow firms to update their priors around which approaches to pursue.
In turn, these updated priors should help firms in the ecosystem mitigate the curse of dimensionality (see Figure \ref{fig3}), and more generally, mitigate the risk of developing a quantum computer.
If, for example, the ability to transparently benchmark leads the ecosystem to pursue two fewer approaches for every layer of the stack, this would lead to $8 \times 8 \times 8 \times 8 \times 8 = 8^5 = 32,768$ remaining configurations that would need to be explored in our numerical example. This reduced number of configurations should again reduce the expected time and cost it will take to create a fault tolerant quantum computer.

\subsection{As a mechanism to improve the supply side of the market for labor}

The next catalyst relates to a frequently cited reason for individuals to voluntarily work on an open-source project. Specifically, individuals may choose to work on an open-source project if it allows them the opportunity to display their talent and elevate their career prospects \cite{lerner}. This incentive may be particularly strong when an individual’s contributions to the open-source project can be properly attributed back to the individual. Separately, as we suggest above, the market for labor in the quantum ecosystem is novel because: 1.) the demand for talent currently outstrips the supply,
2.) there are limited mechanisms for individuals with quantum-adjacent skillsets to demonstrate their ability to the broader ecosystem, 3.) there are limited avenues for individuals in general to experiment with and learn about the inner workings of a quantum computer as a result of the highly proprietary nature of the ecosystem. 

An open-source quantum computing project may help alleviate these issues. For example, access to an open-source quantum computer may allow individuals to potentially upskill their abilities and, if setup correctly, provide a record of any innovations they generate. These outcomes may be particularly important for scientists with quantum-adjacent skill sets, as these stakeholders may have fewer means to signal their ability to prospective employers. It follows that if individuals are able to use an open-source quantum computer to: 1.) demonstrate ability via a publicly available record, and 2.) upskill abilities around quantum, then an open-source quantum project may improve the supply side of the market for labor for the quantum ecosystem as a whole. 

If an open-source quantum project can serve as a mechanism to increase supply in the market for labor, then conventional economics would suggest that it may also make the cost of labor in the market more efficient. In turn, if labor costs become more efficient, then aggregate costs to developing a given approach to a layer of the stack should also become more efficient, and firms may be more able to cost effectively diversify their approaches instead of getting “locked-in” to a specific approach. If a firm could, for example, pursue two approaches at every layer of the stack instead of one, then it could pursue $2 \times 2 \times 2 \times 2 \times 2 = 2^5 = 32$ possible configurations to building a quantum computer instead of one. It follows that an open-source quantum project may help mitigate the curse of dimensionality by making the cost of labor more efficient. Note that this outcome does not arise because the open-source project leads to a reduction in the total possible number of configurations to building a quantum computer. Rather, the curse is mitigated in this instance because it may allow firms to more cost efficiently develop multiple approaches to building a quantum computer. This would be equivalent to adding more possibilities for completed stacks on the right side of Figure~\ref{fig2}.

\section{An Open-Source Project as a Complement to the Proprietary Firms in the Ecosystem}

The introduction of an open-source quantum initiative may lead to improved liquidity in the market for labor, a more transparent mechanism for benchmarking new technologies, and more hardware agnostic offerings. These outcomes may reduce the number of possible configurations to building a quantum computer while also improving firms’ ability to diversify their approaches. Both results should in turn mitigate the curse of dimensionality. Using our original numerical example, rather than pursuing one configuration out of 100,000 possible configurations to build a quantum computer, an open-source project may be the catalyst that allows proprietary quantum computing firms to pursue more than one configuration out of less than 100,000 possible configurations to building a quantum computer. In expectation, such an outcome should shorten the path to building a fault tolerant quantum computer for the proprietary companies in the ecosystem. 

Although an open-source quantum project may eventually compete directly with proprietary quantum computing companies once the technology matures, it may also alleviate market frictions during the R\&D phase of the technology cycle that may make the whole ecosystem better off.  Interestingly, by potentially reducing the time it takes to create a fault-tolerant quantum computer, an open-source quantum project may also be the catalyst that allows initiatives other than its own to win the race to fault tolerance. In this respect, an open-source quantum project may initially provide a complementary role in the market for quantum computers instead of the more competitive role that we typically associate with open-source projects. These collective insights may have parallels for other settings where an open-source project enters into a market where the core technology has yet to be developed and commercialized.

\backmatter

\bmhead{Acknowledgements}
We thank Avi Goldfarb, Simon Cross, and Greg Dick for a number of interesting discussions. We also benefited from discussions in the Creative Destruction Lab (CDL) program for quantum startups. 
RGM is supported by the Natural Sciences and Engineering Research Council of Canada (NSERC) and the Perimeter Institute for Theoretical Physics. Research at Perimeter Institute is supported in part by the Government of Canada through the Department of Innovation, Science and Economic Development Canada and by the Province of Ontario through the Ministry of Colleges and Universities.
This article is dedicated to the memory of CDL Quantum's first Academic Director, Peter Wittek, who was a pioneer in open-source quantum science.








\end{document}